\documentclass[runningheads]{llncs}
\usepackage[T1]{fontenc}
\usepackage{graphicx}
\usepackage{subcaption}
\usepackage{float}
\usepackage{tabularx, makecell, booktabs}
\usepackage{url}
\usepackage{changepage} 
\usepackage{hyperref}  
\usepackage{doi}       
\usepackage{enumitem}
\usepackage{amsmath}
\begin{document}
%





\title{Selective Phase-Aware Training of nnU-Net for Robust Breast Cancer Segmentation in Multi-Center DCE-MRI}
%

\author{Beyza Zayim\inst{1}\orcidID{0009-0002-0054-8466}\thanks{These authors contributed equally. A. Ikram was a former student at University of Algiers 1.} \and
Aissiou Ikram\inst{2}\orcidID{0009-0008-2845-3606}\protect\footnotemark[1] \and
Boukhiar Naima\inst{2}\orcidID{0009-0005-0064-2737}\protect\footnotemark[1]}

\institute{Université Bourgogne Europe, CNRS, ICMUB UMR 6302 Laboratory, Dijon 21000, France \and
University of Algiers 1 Benyoucef Benkhedda Algiers, Algeria}

\maketitle 

\begin{abstract}
Breast cancer remains the most common cancer among women and is a leading cause of female mortality. Dynamic contrast-enhanced MRI (DCE-MRI) is a powerful imaging tool for evaluating breast tumors, yet the field lacks a standardized benchmark for analyzing treatment responses and guiding personalized care.We participated in the MAMA-MIA Challenge’s Primary Tumor Segmentation task and this work presents a proposed a selective, phase-aware training framework for the nnUNet architecture, emphasizing quality-focused data selection to strengthen model robustness and generalization. We employed the No New Net (nnU-Net) framework with a selective training strategy that systematically analyzed the impact of image quality and center-specific variability on segmentation performance. Controlled experiments on the DUKE, NACT, ISPY1, and ISPY2 datasets revealed that including ISPY scans with motion artifacts and reduced contrast impaired segmentation performance, even with advanced preprocessing, such as contrast-limited adaptive histogram equalization (CLAHE). In contrast, training on DUKE and NACT data, which exhibited clearer contrast and fewer motion artifacts despite varying resolutions, with early phase images (0000–0002) provided more stable training conditions. Our results demonstrate the importance of phase-sensitive and quality-aware training strategies in achieving reliable segmentation performance in heterogeneous clinical datasets, highlighting the limitations of the expansion of naive datasets and motivating the need for future automation of quality-based data selection strategies.

\end{abstract}
\section{Introduction}

Breast cancer is the most frequently diagnosed cancer in women and a leading cause of cancer-related mortality worldwide. Dynamic contrast-enhanced magnetic resonance imaging (DCE-MRI) is widely used in clinical practice to evaluate breast tumors because of its ability to visualize vascularization and contrast uptake. Accurate tumor segmentation in DCE-MRI is critical for measuring tumor size, monitoring treatment response, and informing surgical and therapeutic decisions. However, manual segmentation is time-consuming, prone to observer bias, and difficult to scale across large datasets, making automation essential in modern clinical workflows.

A major obstacle in developing robust automatic segmentation models for DCE-MRI is the variability in the imaging protocols across institutions. Multicenter datasets often include images acquired using different scanners, contrast protocols, and spatial resolutions. These multi-center DCE-MRI datasets often display extensive heterogeneity in image quality, resolution, and artifact prevalence as recently detailed in the official MAMA MIA Scientific Data publication \cite{ref_article9}. While ISPY2 represents the largest subset of our dataset (980 cases out of 1506), preliminary analysis revealed significant quality challenges including motion artifacts, reduced contrast, and inconsistent tumor boundaries that negatively impacted model training stability. This motivated our selective training approach, although we acknowledge that this creates a trade-off between dataset diversity and training consistency, which merits further investigation. Our approach focuses on understanding how data quality characteristics affect segmentation performance rather than maximizing dataset size.

In this study, we propose a robust and generalizable segmentation approach for breast tumor detection in DCE-MRI using the nnU-Net framework, emphasizing quality-aware data selection and temporal phase sensitivity.
Our key contribution is a selective phase-aware training strategy for nnU-Net that filters low-quality multi-center DCE-MRI scans and improves segmentation generalization. Through systematic experiments, we analyzed how scan quality, center variability, and multiphase integration affect performance in an area previously underexplored in breast MRI segmentation.
This strategy aims to enhance model reliability across heterogeneous clinical datasets, making it more suitable for real-world deployment in multi-institutional settings.

\section{Related work }
Automated segmentation algorithms can process medical images significantly faster than manual methods, reducing inter-observer variability and providing consistent and reproducible results. The U-Net architecture has been widely adopted because of its skip connections and ability to preserve spatial information, which enhances segmentation accuracy. Building upon this foundation, the nnU-Net framework introduced by Isensee et al. \cite{ref_article1} improves upon U-Net by offering a fully automated, self-configuring deep learning pipeline that requires no manual tuning of hyperparameters. nnU-Net automatically determines whether to use 2D, 3D, or 3D-cascade U-Net architectures and optimizes the preprocessing, training, and postprocessing steps based on the dataset properties. This adaptability renders it a highly generalizable and competitive solution for a wide range of biomedical segmentation tasks. Recent research has also explored the utility of nnU-Net beyond segmentation, utilizing its powerful feature representations for downstream tasks, such as classification, thereby expanding its impact within the medical imaging field.
Several examples illustrate the versatility and effectiveness of nnU-Net in both segmentation and extended application. One such study was conducted by Li et al. \cite{ref_article2}. proposed a multitask learning framework that uses nnU-Net to segment histopathology images and leverages its features for breast cancer classification. Another example is Cheng et al. \cite{ref_article3}. combined nnU-Net-based brain tumor segmentation with radiomics to improve the accuracy of tumor grading. In addition, Tang et al. \cite{ref_article4}. utilized hybrid features extracted via nnU-Net and deep CNNs to classify liver tumors following segmentation. This trend continues in the current MICCAI 2025 Challenge, which evaluates algorithms for primary tumor segmentation and prediction of pathological complete response (pCR) using multi-center breast cancer DCE-MRI data—an integrated task that exemplifies the combined use of segmentation and classification for clinical decision-making.
\section{Methods}
The segmentation pipeline consisted of four main stages: dataset preparation, preprocessing, model training using nnU-Net, and post-processing. Preprocessing was applied to standardize the voxel size and enhance data quality according to the default nnU-Net settings, including resampling, z-score normalization, and cropping. The nnU-Net model was employed using its 3D full-resolution configuration, and 5-fold cross-validation was performed to ensure a robust performance and generalizability. In the post-processing step, small false positives were removed by retaining only the largest connected component to improve the final-segmentation output. The dataset can be accessed via the official challenge page on Synapse \cite{ref_dataset,ref_article9}. 
\vspace{-16pt} 
\begin{figure}

\includegraphics[width=0.9\textwidth]{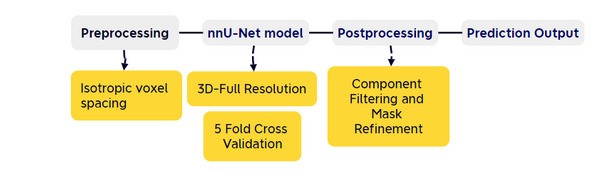}
\vspace{-6pt} 
\caption{Segmentation pipeline overview.} 
\vspace{-16pt} 
\label{fig1}
\end{figure}
\vspace{-10pt} 
\subsection{Dataset}
We used the Synapse dataset released for the MAMA-MIA Challenge \cite{ref_dataset,ref_article9}, as formally described in the official benchmark publication by Garrucho et al.  \cite{ref_article9}, which can be accessed via the official challenge page on Synapse, consisting of 1,506 dynamic contrast-enhanced MRI (DCE-MRI) cases from four centers: DUKE (291), ISPY1 (171), ISPY2 (960), and NACT (64). Each patient scan included between three and six temporal DCE phases and was accompanied by 49 harmonized clinical and demographic variables. Both automated and expert tumor annotations were provided; in all experiments, we used only the expert labels to ensure reliability.
A total of 1,200 cases were allocated for training and 306 for testing in this study. To assess the effects of center variability and phase configuration, we conducted multiple training experiments. We began with 80 DUKE-only cases, then expanded to approximately 400 mixed-center cases, followed by training on the full dataset. All initial experiments used Phase 0002, which was selected for its high tumor-to-background contrast.
We observed that including the ISPY1 and ISPY2 data degraded performance owing to lower image quality, noise, and blurred tumor boundaries. In contrast, DUKE and NACT offered a clearer contrast and more stable training. Therefore, our final setup used only the 247 DUKE and NACT cases, with three temporal phases (0000–0002) as input. This selective, phase-aware strategy improved the segmentation robustness and mitigated the domain shift across the centers. 

\subsection{Preprocessing}

All DCE-MRI volumes were first resampled to isotropic voxel spacing (1 mm³) to ensure consistent resolution across all spatial dimensions. This step promotes anatomical consistency across patients and imaging centers, which is essential for nnU-Net training. Following this, Z-score normalization was applied on a per-volume basis, scaling the voxel intensities to have zero mean and unit variance. These standard preprocessing steps helped mitigate inter-patient and inter-center variability, particularly in the intensity profiles caused by differences in scanners and acquisition protocols.
During the early experiments, we observed that the image quality varied considerably across the centers. Specifically, the ISPY1 and ISPY2 scans often exhibited lower contrast, increased noise, and less distinct tumor boundaries than those of the DUKE and NACT scans. To address this, we tested the application of CLAHE, a local histogram-based enhancement method that is often used to improve the contrast in medical images. While CLAHE initially increased contrast in lower-quality scans ( Fig. 2), it also introduced structured artifacts, such as banding and over-sharpening, particularly in the ISPY 1 and 2 collections of the dataset ( Fig. 2). These distortions negatively affect the segmentation performance and stability during training. 

\vspace{-1em}

\begin{figure}[H]
\centering
\hspace*{-1em}
\includegraphics[width=0.95\textwidth]{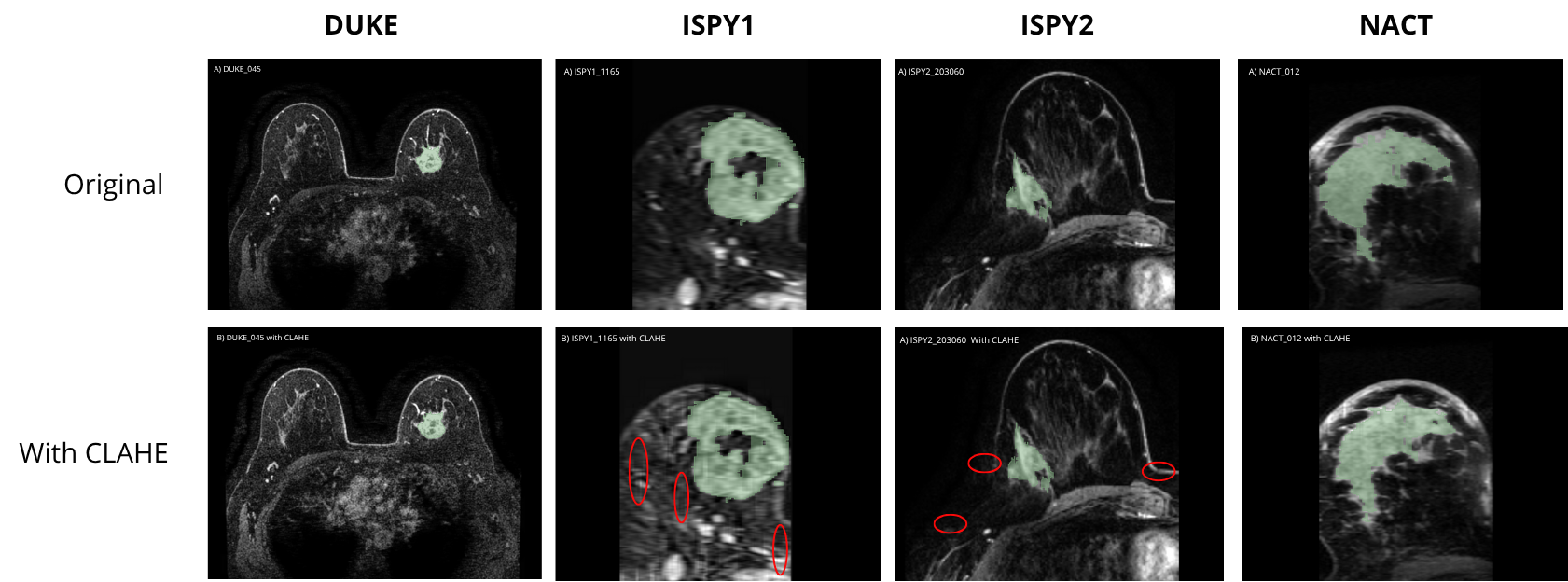}
\caption{DCE-MRI images before (top) and after CLAHE preprocessing (bottom). Red circles show CLAHE-induced artifacts across different centers.}
\label{fig2}
\end{figure}
\vspace{-1em}
\noindent

Consequently, CLAHE was excluded from our final pipeline. We retained only isotropic resampling and Z-score normalization, which proved sufficient for enhancing intensity consistency without introducing artificial features. This streamlined preprocessing configuration provided stable and reproducible results across all experimental conditions while preserving the native contrast characteristics of the original scans\cite{ref_article6} \cite{ref_article7}.
\subsection{Model Architecture and Training Strategy}

The nnU-Net framework, an adaptive deep learning model based on the U-Net architecture, was employed for biomedical image segmentation. It leverages an automated pipeline configuration for preprocessing, network tuning, training, and post-processing.

As part of the preprocessing pipeline, z-score normalization was applied to standardize the image intensities, along with resampling and cropping. We implemented the 3D full-resolution U-Net variant, trained using batch-wise Dice loss, and optimized with the Adam optimizer (learning rate = $1\mathrm{e}{-4}$, weight decay = $3\mathrm{e}{-5}$) on an NVIDIA A100 80GB GPU (PyTorch 2.6.0, CUDA 11.8).

Training was conducted for 1000 epochs (batch size = 2) with early stopping and was augmented using spatial transformations (rotations, scaling, elastic deformations, flipping, cropping) and intensity transformations (gamma correction, noise, blur, contrast adjustments).

During inference, we ensembled the models using 5-fold cross-validation to enhance robustness. To optimize the segmentation performance, we conducted six systematic experiments.
\begin{enumerate}[label=\arabic*., leftmargin=2em]
    \item Our Baseline: 80 DUKE samples, Phase 0002 only
    \item Mixed centers: 400 samples across all centers, Phase 0002 only
    \item Enhanced preprocessing: 400 samples with CLAHE, Phase 0002 only
    \item Full dataset: 1200 samples across all centers, Phase 0002 only
    \item Quality-filtered: Combined DUKE and NACT data, Phase 0002 only
    \item Multi-phase: DUKE and NACT data, Phases 0000--0002
\end{enumerate}

The best model was selected based on the validation results from the MAMA-MIA Challenge, balancing generalization and accuracy across datasets.

\subsection{Evaluation Metrics}

To evaluate the performance of the deep learning model for tumor segmentation, we use three key metrics: Dice Score, Hausdorff Distance, and Fairness Score.

\textbf{Dice Score} is a widely used overlap-based metric in medical image segmentation that quantifies the similarity between the predicted segmentation and the ground truth. It is defined as:
\begin{equation}
\text{Dice} = \frac{2 \times |A \cap B|}{|A| + |B|},
\end{equation}
where $A$ represents the set of predicted tumor pixels, and $B$ is the set of ground truth tumor pixels. A Dice score of 1 indicated a perfect overlap, whereas a score of 0 indicated no overlap. This metric is particularly useful for evaluating the accuracy of the model in capturing the tumor region.

\textbf{Hausdorff Distance} evaluates the accuracy of the predicted tumor boundary by measuring the maximum distance between the boundary points of the predicted and ground truth segmentations. It is defined as:
\begin{equation}
H(A, B) = \max(h(A, B), h(B, A)),
\end{equation}
where $h(A, B) = \max_{a \in A} \min_{b \in B} \|a - b\|$. This metric captures the worst-case discrepancies between the two contours. A lower Hausdorff Distance indicates better boundary alignment and is especially important when precision at the tumor margins is critical.

\textbf{Fairness Score} is used to assess whether the segmentation model performs equitably across different subgroups, such as patient demographics (e.g., gender or age groups). One common fairness measure is Demographic Parity, given by
\begin{equation}
\text{DP} = \left| P(\hat{Y} = 1 \mid G = \text{group}_1) - P(\hat{Y} = 1 \mid G = \text{group}_2) \right|,
\end{equation}
where $\hat{Y}$ is the model prediction and $G$ represents the group variable. A fairness score close to zero indicates that the model performance is balanced across groups, helping to ensure that no subgroup is disproportionately under- or over-segmented.
\subsection{Postprocessing}
To refine the segmentation output and reduce the number of false positives, we applied a post-processing step that retained only the largest connected component in each predicted mask. This approach is based on the clinical observation that the primary tumor is typically the largest contiguous lesion in the breast region. We performed 3D connected component analysis using the \texttt{measure.label()}function from the \texttt{scikit-image} library, which labels the connected regions in an integer array. The regions were then filtered based on size, and only the largest component was retained. The resulting mask was resampled to the original image space and stored as the final prediction.

This simple yet effective strategy improves the segmentation quality by eliminating small scattered regions caused by noise or model uncertainty, particularly in low-contrast scans. This led to higher Dice scores and more anatomically plausible outputs without requiring any changes to the model architecture \cite{ref_article7} \cite{ref_article8}.

\section{Simulation results}

Our experiments demonstrated varying segmentation performances across different training strategies, as measured by the Dice scores (see Table~\ref{tab:results}). When using only Phase 0002 data, the model achieved consistently high performance on the DUKE-001 dataset (Dice scores: 0.8625--0.8914), while showing more variable results on ISPY datasets (0.6739--0.7901 for \texttt{ISPY1\_113}; 0.5227--0.8426 for \texttt{ISPY2\_332}).

Notably, expanding the training data from 80 to 400 samples showed minimal improvement, and further increasing to 1200 samples actually degraded the performance on some test sets. This suggests that simply adding more data does not necessarily guarantee better generalization performance.

The application of CLAHE preprocessing yielded mixed effects: it substantially improved performance on the \texttt{NACT\_64} dataset (Dice scores $\sim$0.95) while providing no clear benefit for other datasets.

Interestingly, the combination of DUKE and NACT data with 5-fold cross-validation achieved the best overall validation performance (Dice = 0.62), outperforming other single-phase approaches. The multi-phase experiment (Phases 0000-0002) achieved the highest validation performance of 0.72, indicating that temporal information integration represents a promising direction. Official test results from the challenge organizers are pending following our submission. (see Table~\ref{tab:results}).

These findings highlight the complex relationship between data quantity, preprocessing methods, and domain-specific characteristics in achieving a robust segmentation performance.
\vspace{-10pt} 
\begin{table}[H]  
\centering
\resizebox{\textwidth}{!}{%
\begin{tabular}{l|cccccc}
\hline
\textbf{Experiments} & 1 & 2 & 3 & 4 & 5 & \textbf{6} \\
\textbf{Phase} & (\textit{Phase 0002} only) & (\textit{Phase 0002} only) & (\textit{Phase 0002} only) & (\textit{Phase 0002} only) & (\textit{Phase 0002} only) & \textbf{(Phases 0000-0002)} \\
\hline
\textbf{Data} & DUKE-80 & 400-All centers & 400-All centers-CLAHE & 1200-All data & DUKE+NACT & \textbf{DUKE+NACT} \\
\textbf{Fold} & 5 folds & 5 folds & 5 folds & 1 fold & 5 folds & \textbf{5 folds} \\
\textbf{DUKE\_001 (291 data)} & 0.8914 & 0.8859 & 0.8879 & 0.8894 & 0.8625 & \textbf{0.9394} \\
\textbf{ISPY1\_1183 (171 data)} & 0.7901 & 0.7474 & 0.7246 & 0.6739 & 0.7196 & \textbf{0.7640} \\
\textbf{ISPY2\_332655 (980 data)} & 0.8426 & 0.8420 & 0.8243 & 0.5227 & 0.8111 & \textbf{0.8967} \\
\textbf{NACT\_64 (64 data)} & 0.7357 & 0.9489 & 0.9486 & 0.9334 & 0.9514 & \textbf{0.9580} \\
\textbf{Validation Performance} & 0.59 & 0.55 & 0.49 & 0.45 & 0.62 & \textbf{0.72} \\
\hline
\end{tabular}%
}
\vspace{5pt}
\caption{Dice scores based on training strategies}
\vspace{-16pt} 
\label{tab:results}
\end{table}
\vspace{-30pt}
\begin{figure}[H]
\centering
\hspace*{-0.05\textwidth} 
\includegraphics[width=1.1\textwidth]{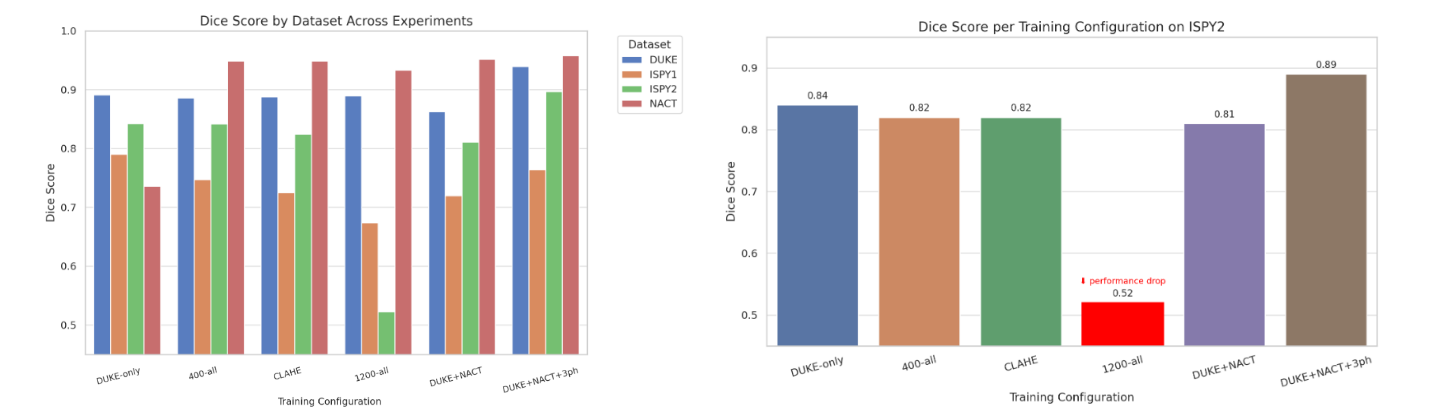}
\caption{Dice scores across experimental configurations. Left: Performance on all the datasets. Right: ISPY2-specific results showing consistent performance patterns.}
\label{graph}
\end{figure}

\vspace{-16pt} 

We acknowledge that our selective training approach did not surpass the MAMA-MIA challenge baseline performance on the validation set. Our primary objective was to systematically investigate the relationship between data quality, center variability, and model robustness, rather than optimizing for maximum validation scores. These results demonstrate that simply increasing the training data volume does not guarantee improved performance when the data quality varies significantly across centers. The modest improvement from 0.59 (DUKE-80) to 0.62 (DUKE+NACT) suggests that our quality-focused selection strategy provides stable but incremental benefits, while the multi-phase approach (0.72) shows greater promise for future development. 

Figure 3 provides a comprehensive overview of our experimental progression and decision-making process. The left panel illustrates the performance trajectory across all six experimental configurations, revealing several key insights regarding the performance trajectory. First, expanding from DUKE-only (80 samples) to mixed-center data (400 samples) showed minimal improvement, suggesting that simply increasing the dataset size does not guarantee better results. Second, the application of CLAHE preprocessing degraded performance across most datasets, confirming our decision to exclude it from the final pipeline. Third, the dramatic performance drop when using the full 1200-sample dataset demonstrates that including lower-quality data can harm the model performance, even when combined with high-quality samples. The right panel focuses specifically on the ISPY2 performance patterns, showing consistent trends that mirror the overall dataset behavior. Most notably, the multi-phase approach (DUKE+NACT with Phases 0000-0002) achieved the highest validation performance of 0.72, indicating that temporal information integration represents a promising direction for future development.

\section{Discussions}
The application of isotropic spacing in MRI data preprocessing increases the data size, introducing challenges in storage and computational management. However, experiments have shown that simply expanding the dataset does not consistently improve the model performance. Instead, selecting data based on MRI quality, particularly high-resolution, artifact-free scans from different centers, proved critical for reliable outcomes, underscoring the impact of image quality on model generalization. These findings highlight the delicate balance between data volume, uniformity, and quality in medical imaging analysis, necessitating the development of optimized preprocessing strategies to maintain accuracy while managing computational costs.

The best validation performance was considered when selecting the data. Because the best validation performance was achieved with the DUKE and NACT datasets, these datasets were used for the next experimental strategies, such as adding more than one phase.
Interestingly, incorporating multiple MRI phases alongside a larger dataset enhanced generalization, suggesting that diverse yet uniformly processed data can improve model robustness. However, given the variability in scan quality, future studies should explore quality-aware selective sampling within the nnU-Net pipeline. By automatically prioritizing diagnostically relevant scans (e.g., excluding motion artifacts or low-resolution images), the model can be trained on a smaller, high-quality subset without sacrificing performance. Future experiments could compare random sampling with selective approaches (e.g., quality-based filtering, diversity-aware selection, or uncertainty-driven sampling) to determine whether a curated dataset achieves results comparable to the full isotropic-spaced dataset while reducing computational overhead. This approach may optimize the trade-off between efficiency and accuracy in nnU-Net-based segmentation tasks. 

This study has several limitations that should be acknowledged. First, our selective approach excludes a substantial portion of the available training data (ISPY1 and ISPY2), which may limit generalizability to real-world clinical scenarios in which such quality filtering is not feasible. Second, our quality assessment was performed through visual inspection combined with quantitative metrics (Dice, Fairness, NormHD), which introduced subjectivity despite the systematic approach. Third, our contribution is primarily empirical rather than methodological; we provide a systematic analysis of data-centric factors affecting segmentation performance rather than novel architectural innovations. Finally, while our fairness analysis showed comparable demographic balance, the reduced dataset diversity may impact performance across different patient subgroups in broader clinical deployment.
\section{Conclusion}
This study provides empirical insights into data quality effects on multi-center breast DCE-MRI segmentation. Although our selective training approach did not surpass the challenge baselines, it demonstrated that dataset curation and quality assessment are as important as architectural innovations for robust clinical AI systems.
Our findings reveal a fundamental trade-off between dataset diversity and quality consistency in multicenter studies. The promising multi-phase results (0.72 validation performance) suggest that temporal information integration warrants further investigation. Future studies should focus on automated quality assessment methods and domain adaptation strategies that can leverage the full dataset diversity while maintaining reliability.
By documenting these data-centric factors, we contribute practical insights for training data preparation in clinical AI applications with significant cross-center variability.

\paragraph{Acknowledgments}

This work was supported by the MAMA-MIA Challenge 2025, which provided a foundational dataset and framework for advancing generalizability and fairness in breast MRI tumor segmentation and treatment response prediction. We extend our gratitude to the challenge organizers, data contributors, and annotators for their efforts in curating these valuable resources. We also acknowledge the computational resources provided by the Université de Bourgogne, which enabled the training and evaluation of the nnU-Net models. Finally, we thank our collaborators for their insightful feedback and the broader medical imaging community for their open-source tools that made this study possible.

\paragraph{Disclosure of Interests.}The authors declare no conflicts of interest related to this study. This research was conducted independently of any commercial or financial relationships that could be construed as potential conflicts of interest. Participation in the MAMA-MIA Challenge 2025 was solely for scientific and clinical advancement of breast MRI segmentation and treatment response prediction.

%
%
%

\begin{thebibliography}{11}
\bibitem{ref_article1} Isensee, F., Jaeger, P.F., Kohl, S.A.A., Petersen, J., Maier-Hein, K.H.: nnU-Net: A self-configuring method for deep learning-based biomedical image segmentation. \emph{Nature Methods} \textbf{18}, 203--211 (2020). \doi{10.1038/s41592-020-01008-z}

\bibitem{ref_article2} Li, Y., Liu, W., Li, Z., Zhang, Y., Zhou, J., Liu, Y.: An nnU-Net-based framework for breast cancer histopathological image classification using multi-task learning. \emph{Frontiers in Oncology} \textbf{12}, 837424 (2022). \doi{10.3389/fonc.2022.837424}

\bibitem{ref_article3} Cheng, Y., Wang, S., Lin, H., Zhang, Y., Zhang, Q.: Integrating nnU-Net-based tumor segmentation with radiomics for brain tumor grading. \emph{IEEE Access} \textbf{11}, 32005--32014 (2023). \doi{10.1109/ACCESS.2023.3252876}

\bibitem{ref_article4} Tang, Y., Wang, Z., Li, Q., Liu, H., Zhao, Y.: Liver tumor classification using hybrid features extracted by nnU-Net segmentation and deep CNNs. \emph{Computers in Biology and Medicine} \textbf{138}, 104923 (2021). \doi{10.1016/j.compbiomed.2021.104923}

\bibitem{ref_article5}
Schwarzhans, F., George, G., Sanchez, L.E., Zaric, O., Abraham, J.E., Woitek, R., Hatamikia, S.: Intensity Normalization Techniques and Their Effect on the Robustness and Predictive Power of Breast MRI Radiomics. \emph{arXiv} (2024). \url{https://arxiv.org/abs/2406.01736}

\bibitem{ref_article6}
Nyúl, L.G., Udupa, J.K.: On standardizing the MR image intensity scale. \emph{Magnetic Resonance in Medicine} \textbf{42}(6), 1072--1081 (1999). \doi{10.1002/(sici)1522-2594(199912)42:6}
\bibitem{ref_article7}
Kearney, V., Haaf, S., Sudhyadhom, A., Valdes, G., Solberg, T.: An open-source tool for automatic segmentation of the prostate on MRI. \emph{Medical Imaging in Radiation Oncology} \textbf{1}, 10--17 (2021). \doi{10.1016/j.miro.2021.04.002}

\bibitem{ref_article8}
Taha, A.A., Hanbury, A.: Metrics for evaluating 3D medical image segmentation: analysis, selection, and tool. \emph{Nature Methods} \textbf{18}, 1139--1147 (2021). \doi{10.1038/s41592-020-01008-z}

\bibitem{ref_github}
Zayim, B. MICCAI Challenge Code Repository. GitHub (2025). \url{https://github.com/beyza17/MICCAI_Challange}

\bibitem{ref_dataset}
MAMA-MIA Challenge Dataset. Synapse (2025). \url{https://www.synapse.org/Synapse:syn60868042/wiki/628716}


\bibitem{ref_article9}
L. Garrucho, K. Kushibar, C.-A. Reidel, S. Joshi, R. Osuala, A. Tsirikoglou, M. Bobowicz, J. del Riego, A. Catanese, K. Gwoździewicz, M.-L. Cosaka, P. M. Abo-Elhoda, S. W. Tantawy, S. S. Sakrana, N. O. Shawky-Abdelfatah, A. M. A. Salem, A. Kozana, E. Divjak, G. Ivanac, K. Nikiforaki, M. E. Klontzas, R. García-Dosdá, M. Gulsun-Akpinar, O. Lafcı, R. Mann, C. Martín-Isla, F. Prior, K. Marias, M. P. A. Starmans, F. Strand, O. Díaz, L. Igual, and K. Lekadir, ``A large-scale multicenter breast cancer DCE-MRI benchmark dataset with expert segmentations,'' \emph{Scientific Data}, vol. 12, no. 1, p. 453, 2025, doi: \doi{10.1038/s41597-025-04707-4}


\end{thebibliography}
%

\end{document}